\documentclass[12pt]{elsart}
\usepackage{epsfig}
\usepackage[usenames]{color}
\journal{ }

\def\epj#1#2#3#4{    {Eur. Phys. J. }#1 {\bf #2} (#3) #4}
\def\np#1#2#3#4{    {Nucl. Phys. }#1 {\bf #2} (#3) #4}
\def\npp#1#2#3{    {Nucl. Phys. Proc. Suppl. }{\bf #1} (#2) #3}
\def\pl#1#2#3#4{    {Phys. Lett. }#1 {\bf #2} (#3) #4}

\def\pr#1#2#3#4{    {Phys. Rev. }#1 {\bf #2} (#3) #4}

\def\prl#1#2#3{    {Phys. Rev. Lett. }{\bf #1} (#2) #3}

\def\mpl#1#2#3#4{   {Mod. Phys. Lett. }#1 {\bf #2} (#3) #4}

\def\re#1{{\mathrm Re}\left\{#1\right\} }
\def\im#1{{\mathrm Im}\left\{#1\right\} }
\newcommand{\com}[1]{ \par }
\def\spa{\hspace*{-.65cm}}

\def\evg{\, g_{eV}^\gamma}
\def\tvg{\, g_{\tau V}^\gamma}
\def\evz{\, g_{eV}^Z}
\def\eaz{\, g_{eA}^Z}
\def\tvz{\, g_{\tau V}^Z}
\def\taz{\, g_{\tau A}^Z}
\def\tz{\, g_{\tau T}^Z}
\def\tg{\, g_{\tau T}^\gamma}
\def\evg2{(g_{eV}^\gamma)^2}
\def\tvg2{(g_{\tau V}^\gamma)^2}
\def\evz2{(g_{eV}^Z)^2}
\def\eaz2{(g_{eA}^Z)^2}
\def\tvz2{(g_{\tau V}^Z)^2}
\def\taz2{(g_{\tau A}^Z)^2}
\def\tz2{(g_{\tau T}^Z)^2}
\def\tg2{(g_{\tau T}^\gamma)^2}

\def\re#1{{\mathrm Re}\left\{#1\right\} }
\def\im#1{{\mathrm Im}\left\{#1\right\} }

\def\dps{\displaystyle}
\def\spa{\hspace*{-.65cm}}
\newcommand{\beq}{\begin{equation}}
\newcommand{\eeq}{\end{equation}}
\newcommand{\bi}{\begin{itemize}}
\newcommand{\ei}{\end{itemize}}
\newcommand{\bea}{\begin{eqnarray}}
\newcommand{\eea}{\end{eqnarray}}
\newcommand{\bes}{\begin{eqnarray*}}
\newcommand{\ees}{\end{eqnarray*}}

\begin{document}
\begin{flushright}FTUV-08-1507\end{flushright}
\begin{frontmatter}
\title{Tau spin correlations and the anomalous magnetic moment}
\author[Valencia]{J. Bernab\'eu},
\author[Montevideo]{G. A. Gonz\'alez-Sprinberg}
\author[Valencia]{J. Vidal}
\address[Valencia]{Departament de F\'{\i}sica Te\`orica
Universitat de Val\`encia, E-46100 Burjassot,Val\`encia, Spain\\
and\\
IFIC, Centre Mixt Universitat de Val\`encia-CSIC, Val\`encia, Spain}
\address[Montevideo]{Instituto de F\'{\i}sica,
 Facultad de Ciencias, Universidad de la Rep\'ublica,
 Igu\'a 4225, 11400 Montevideo, Uruguay}

\journal{Nuclear Physics B}
  
\begin{abstract}

We show that  the precise determination  of the Tau magnetic properties
is possible in the next generation accelerators, specially at B/Flavour factories.
We define spin correlation observables suitable to extract the real part of the magnetic
form factor that, for the first time, will allow to test the standard model-QED 
predictions. In particular, the
predicted QED-dependence with both the momentum transfer and the lepton mass can be
precisely measured. Until now, 
the most stringent bounds on the $\tau$ magnetic moment $a_\tau$ come from LEP data with 
strong assumptions on the physics involved on the observed process. 
In this paper, we find three different combinations of spin correlations of the outgoing Taus that disentangle the magnetic moment form factor of the Tau lepton in the electromagnetic vertex. These combinations of asymmetries also get rid off the contributions coming from Z-mediating amplitudes to the defined correlations. Using unpolarized electron beams and an integrated  luminosity of $15 \times
10^{18} b^{-1}$, the sensitivity to the $\tau$ magnetic moment form factor is of the order $10^{-6}$. This 
sensitivity is two orders of magnitude  better than the present existing high- or low-energy bounds  
on the magnetic moment and would allow its actual measurement with the precision of a few per cent.
\end{abstract}
\end{frontmatter}

\section{Introduction}
The fact that the $\tau$ magnetic properties cannot be investigated in the way as done for a long 
lived particle, where the interaction with an external magnetic field can be directly measured, makes 
more subtle and difficult its determination. As stated in the PDG, the experimental 
determination of the $\tau$ magnetic moment, by the DELPHI Collaboration~\cite{exp}, was done using 
LEP2 data for the $e^+ e^- \rightarrow e^+  e^- \tau^+  \tau^-$ total cross-section, assuming that 
any deviation from the tree level SM prediction was exclusively due to the magnetic anomaly~\cite{cornet}, 
$a_\tau=(g-2)/2$. This indirect measurement bounds the $\tau$ magnetic moment to the values \cite{pdg}:
\beq
 - 0.052 \,< \, a_\tau \,  <\, 0.013 \,\,\;(95\,\%\, C.L.)
\label{mu}
\eeq
In fact, with such assumptions, in some of the quoted experiments the bound that is effectively set 
is the contribution of new physics to the magnetic moment. This is the point  of  view adopted  
in~\cite{arcadi},  where  the most  stringent model-independent limit for the  magnetic properties is obtained: 
\beq
- 0.007  \,  <  \,  a_\tau^{{\rm   New  Phys.}}   \,  <  \,  0.005\,\,
(\,95\,\,\%\,\, C.L.)  
\eeq

The bound given in Eq.(\ref{mu}) is well above the SM prediction \cite{passera}
\beq
a_\tau^{SM}=1177.21(5)\; \times\; 10^{-6},
\eeq
where higher-order QED, hadronic and weak corrections are included and contribute only
to the last figures of this value. In fact, present 
experimental bounds on the $\tau$ lepton magnetic moment anomaly are still more than one 
order of magnitude bigger than the result obtained by Schwinger \cite{sch} at one-loop in QED:
\beq 
a_e=a_\tau= \frac{\alpha}{2 \pi} \simeq 0.00116 \, .
\label{schw}
\eeq

Super B/Flavour factories will produce in the future  about $10^{12}$ Tau pairs \cite{superb}, so that 
high precision measurements of the poorly know properties of the Tau lepton will be 
possible. In this paper we present a set of new observables appropriate to measure the 
magnetic moment form factor of the $\tau$ by analyzing the angular distribution of their 
decay products in $e^+\; e^-$ collisions. For unpolarized electron beams, the imaginary part of the magnetic moment can be observed by measuring the normal polarization of a single Tau \cite{nos08}, whereas
we show here that only spin correlations of the outgoing Taus are sensible to the real 
part of the magnetic moment. For polarized electron beams, we have shown in Ref.~\cite{nos08} that the real part of the magnetic moment can be observed by measuring the 
transverse and longitudinal polarization of a single Tau. We find that a sensitivity of 
the order of $10^{-6}$ can be 
achieved using high statistics facilities for Tau pair production on the top of the 
$\Upsilon$ resonance. Comparable sensitivities were found in Ref.~\cite{nos08} by using polarized 
electron beams and observables built on the polarization analysis of a single Tau.

\section{Magnetic moment form factor}

The most general Lorentz invariant structure describing the
interaction of a vector boson $V$  with two on-shell fermions $f\bar{f}$ can be
written in terms of the conserved current $J^\mu$ as: 
\bea
&&\spa \langle f(p_-)\bar{f}(p_+)|\,J^\mu(0)\,|0\rangle= \nonumber\\
&& e\, \bar{u}(p_-) \left[  \gamma^\mu\; F_1
+\frac{(i\, F_2+F_3\gamma_5)}{2m_f}\sigma^{\mu\nu}q_\nu
+\left(q^2\gamma^\mu-q^\mu\not{\! q}\right)\gamma_5 F_A\right]v(p_+)
\label{eq:2}
\eea
where $F_1$ is the Dirac form factor ($F_1(0)=1)$,
$F_A$ is the anapole moment, whereas $F_2$ and
$F_3$ parametrize the magnetic and electric dipole moments, respectively.

The determination of the $CP$- violating electric dipole moment, in Super B/Flavour 
factories, has
been studied in detail in \cite{Bernabeu:2004ww,nos07} with similar techniques to the ones 
presented here. The $P$-odd, $T$-even anapole  moment differs from zero due
to weak virtual corrections so that its contribution  will be suppressed
by factors of $q^2/M_Z^2$ compared  to the leading QED corrections considered in this 
paper. The induced magnetic moment form factor $F_2$ is a chirality flip observable 
in the vector current and its determination is the subject of this paper.

Strictly  speaking, the  magnetic moment  anomaly 
\beq
a_f\equiv F_2(q^2=0)
\eeq
is defined with all three fields  entering into the interaction vertex on
their   mass-shell. In Super  B factories the squared center-of-mass energy
$s = q^2$ is of the order $(10\, GeV)^2$ and,  therefore, $F_2(q^2)$ is no longer the 
magnetic anomaly. Note that if the evolution scale is well
above $q^2$, as it is the case for new physics, then $F_2(q^2) \simeq F_2(q^2=0)$ (see Ref.~\cite{arcadi}) but
in the case of interest, for QED, the evolution scale of the form factor is $m_\tau^2 < q^2$ and
the actual value of the form factor  for B-Factories is quite different from the 
magnetic moment anomaly. The
direct computation of the magnetic part of the standard one-loop QED vertex yields
\bea
&&F_2(s)=\left(\frac{\alpha}{2\pi}\right)\frac{2m_\tau^2}{s}\frac{1}{\beta}\left
(\log\frac{1+\beta}{1-\beta}-i\,\pi\right),\quad\mbox{for}
\quad q^2 = s > 4 m_\tau^2,\label{MMff}
\eea
where $\alpha$ is the fine structure constant and $\beta = \left(1-4m_\tau^2 /
s\right)^{1/2}$ is the velocity of the $\tau$.

The $F_2(s)$ form factor is gauge invariant, despite being
an off-shell amplitude. This is so due to the specific gauge-cancellation 
mechanism that operates only in an abelian theory such as QED: the direct box cancels against the crossed one, the 
vector proper vertex against self-energy fermions graphs so the  magnetic moment 
form factor, proportional to $\sigma_{\mu\nu}q^{\nu}$, must be 
separately gauge-independent.

In our case, for $q^2\sim M_\Upsilon^2 \sim (10\, {\rm GeV})^2$,
\beq
F_2(M_\Upsilon^2)=(2.65 - 2.45\, i)\times 10^{-4}.\label{f2}
\eeq

This equation shows that, at this energy, the real and imaginary parts are of the same order and, due to the scale and flavour dependence, 
about $1/4$ of the on-shell magnetic moment. This fact gives us the opportunity 
to see the behaviour of the form factor with the momentum $q^2$, for a pure abelian 
theory as QED, together with its strong dependence on the flavour (mass) of the 
fermion. Note that for other lighter fermions (as the electron) the magnetic moment form factor is vanishing small at the $M_\Upsilon$-energy.
 
For the extraction of $F_2$, in order to eliminate contamination from the box 
diagrams to the measured amplitude, we perform our analysis on top of the Upsilon 
resonance, so that kinematics makes the non-resonant box contributions negligible.

\section{$e^+e^-\longrightarrow \tau^+\tau^-$ at Super B
Factories.}\label{section:gamma}
\begin{figure}[hbtp]
\begin{center}
\epsfig{file=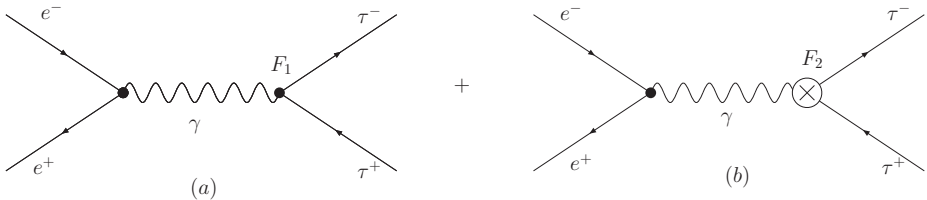,width=.9\textwidth}
\end{center}
\caption{Diagrams: (a) direct $\gamma$ exchange, (b) $F_2$ in $\gamma$ exchange.
}
\label{fig:figura1}
\end{figure}

In  this section  we first  consider the  $\tau$-pair production  in 
$e^+e^-$ collisions
through   direct  $\gamma$   exchange   (diagrams  (a)   and  (b)   in
Fig. \ref{fig:figura1}). Next, we will  show that the basic results of
this section still hold for resonant $\Upsilon$ production.

The differential cross section for $\tau$ pair production, with spin
$\vec{s}_+,\, \vec{s}_ -$, is:

\begin{equation}
\frac{d \sigma(\vec{s}_+,\vec{s}_-)}{d \Omega_{\tau^-}}=
\frac{d \sigma^{0}(\vec{s}_+,\vec{s}_-)}{d  \Omega_{\tau^-}}
+\frac{d \sigma^{S}(\vec{s}_+,\vec{s}_-)}{d  \Omega_{\tau^-}}+\frac{d
\sigma^{SS}(\vec{s}_+,\vec{s}_-)}{d 
\Omega_{\tau^-}}
\label{cross1}
\end{equation}

The first term of Eq.(\ref{cross1}) represents the $\tau$ spin-independent
differential cross section
\begin{equation}
\frac{d \sigma^0(\vec{s}_+,\vec{s}_-)}{d 
\cos\theta_{\tau^-}}= \frac{\pi \alpha^2}{8
s}\beta\left[(2-\beta^2\sin^2\theta_{\tau^-})|F_1(s)|^2+4\re{F_1(s)F_2(s)^*}
\right]
\label{cross01}
\end{equation}
where $\alpha$ is the fine structure constant, the squared center of mass energy
 $s=q^2$ is also the square of the 4-momentum carried by the photon,
and $\gamma=\sqrt{s}/(2 m_\tau)$,
$\beta=\sqrt{1-1/\gamma^2}$,
are the dilation factor and $\tau$ velocity, respectively. 
Note that, at one loop, we have the identity
\beq
\re{F_1(s)F_2(s)^*}=\re{F_2(s)}\,.
\eeq
The $\theta_{\tau^-}$ angle is determined  in the center-of-mass (CM) frame
by the  outgoing $\tau^-$  and the incoming  $e^-$ momenta. In B/Flavour 
factories, a  precise  measurement   of  the
$\theta_{\tau^-}$ angle requires that the  $\tau$ production plane and direction 
of flight are fully reconstructed. In Ref.~\cite{kuhn}  it is shown that this can be
achieved if  both $\tau$'s decay semileptonically.

The second term $\displaystyle \frac{d\sigma^{S}}{d\Omega_{\tau^-}}$ involves 
linear terms in the spin of the $taus$,
\begin{equation}
\frac{d \sigma^{S}(\vec{s}_+,\vec{s}_-)}{d
\cos_{\tau^-}}=\frac{\pi \alpha^2}{4\ s}\; \beta
\; (s_-+s_+)_y\: Y_+\label{cross2}
\end{equation}
where $Y_+$, up to the considered order, is given by 
\beq
Y_+=\gamma\: \beta^2\left(\cos \theta_{\tau^-}\, \sin\theta_{\tau^-}\right)\,
\im{F_2(s)}, \label{cross3}
\eeq
Eq.(\ref{cross3}) shows that the contribution of
the chirality flipping $F_2$ to the normal polarization
is enhanced by a factor of $\gamma$ with respect to
other possible non-chirality flipping contributions. This observable is a 
$P$-even, $T$-odd 
quantity, as corresponds to the interference $\im{F_1(s)\, F_2(s)^*}$.

We work in the center of mass (CM)  frame of reference, and the orientation 
of the coordinate
system is the same as in Ref.~\cite{Bernabeu:2004ww}.
The $\mbox{\boldmath $s$}_\pm$  vectors are the
 $\tau^\pm$  spin vectors in the $\tau^\pm$ rest system, $s_\pm=(0, s^x_\pm,
s^y_\pm, s^z_\pm)$. Polarizations along the directions $x,y,z$ correspond
 to what is called transverse, normal, and longitudinal polarizations,
respectively.

As shown in Eq.(\ref{cross3}), the normal polarization of the outgoing Tau has 
sensitivity 
to the imaginary part of the magnetic moment form factor. In absence of 
$P$-violation, the other components
of a single $\tau$ polarization cannot appear. The polarization of the final 
fermion ($\tau^\pm$) can be studied
by looking at the angular distribution of its decay products \cite{nos08}. 

The last term of Eq.(\ref{cross01}) is proportional to the product of the spins of both $\tau$'s:
\begin{eqnarray}
\frac{d\sigma^{SS}(\vec{s}_+,\vec{s}_-)}{d \cos\theta_{\tau^-}}&=& 
\frac{\pi \alpha^2}{8 s} \beta 
 \left( s_+^x s_-^x C_{xx} + s_+^y s_-^y C_{yy} +
s_+^z s_-^z C_{zz}\right. \nonumber\\
&&\left.\hspace*{4cm}+ (s_+^x s_-^z + s_+^z s_-^x) C_{xz}^+
\right)\label{csection}\end{eqnarray}
where
\bea
C_{xx} &=&
\left((2-\beta^2)\; |F_1|^2 +4\re{F_2}\right)\, \sin^2\theta_{\tau^-}\\
C_{yy} &=& -|F_1|^2\, \beta^2\, \sin^2\theta_{\tau^-}\\
C_{zz}&=&
|F_1|^2(2\cos^2\theta_{\tau^-}+\beta^2\sin^2\theta_{\tau^-})+4\re{F_2}
\cos^2\theta_{\tau^-}\\
C_{xz}^+ &=& \left(|F_1|^2+\gamma^2 (2-\beta^2)\re{F_2}\right)\, 
\frac{1}{\gamma}\, \sin(2 \theta_{\tau^-})
\label{eq:xz}
\eea
%
Equation (\ref{csection}) shows that, as expected in absence of any source of 
$P$-violation, only $P$-even correlations can contribute to the spin-spin differential 
cross section. All the terms are also $T$-even, so only $\re{F_2}$ appears. The $C_{xy}$ 
and $C_{yz}$ correlation terms, being $P$-odd, cannot appear in the given cross section.

The $C_{xz}$ term is the only chirality flipping correlation of the process
and then it has the $F_2$ contribution enhanced by the dilation factor
 $\gamma$, whereas the $|F_1|^2$ form factor, which is chirality conserving,
is suppressed by the Tau mass ($1/\gamma$ factor).

Following Ref.~\cite{tsai}, the differential cross-section for the process 
\[
e^+e^-\rightarrow h^+\bar{\nu}_\tau\,  h^-\nu_\tau
\]
can now be obtained by {\it convoluting} the previous 
$d\sigma\left( e^+e^-\stackrel{\gamma}{\rightarrow} \tau^+(\vec{s}_+)\: 
\tau^-(\vec{s}_-)\right)$, differential cross section, with the  Tau decay 
branching ratios, to get
\begin{eqnarray}
d \sigma\left( e^+e^-
\stackrel{\tau^+\tau^-}{\rightarrow} h^+\bar{\nu}_\tau\, h^-\nu_\tau\right)
&=&4\, d\sigma(\vec{n}^*_+,\vec{n}^*_-)\;
\frac{d\Omega_{h^+}}{4\pi}\frac{d\Omega_{h^-}}{4\pi}\nonumber\\
 &&\times\, Br(\tau^-
\rightarrow h^-\nu_\tau)\;Br(\tau^+
\rightarrow
h^+\bar{\nu}_\tau)
\label{eq:cros01}
\end{eqnarray}
where the spins $\vec{s}_\pm$ in the differential cross-section (\ref{cross1})  
have been substituted by 
\beq \overrightarrow{n}_\pm^*= \mp\alpha_\pm
\frac{\overrightarrow{q}^{  *}_\pm}{
\arrowvert\overrightarrow{q}^{  *}_\pm\arrowvert} =
\mp\alpha_\pm(\sin\theta_{\pm}^*\, \cos\phi_\pm,
\sin\theta_{\pm}^*\, \sin\phi_\pm,\cos\theta_{\pm}^*)\,.\nonumber\\
\eeq
which are proportional to the momenta of the hadrons ($h^\pm$) with moduli  fixed to the
polarization analyzer
\beq
 \alpha_h\equiv \frac{m_\tau^2-2m_h^2}{m_\tau^2+m_h^2}.
 \eeq
The $\phi_{\pm}$ and $\theta^*_{\pm}$ angles are the azimuthal and polar angles
of the
produced hadrons $h^\pm$ ($\hat{q}^*_{\pm}$) in the $\tau^\pm$
rest frame (the * means that the quantity is
given in the $\tau$ rest frame).
 From Eq.(\ref{eq:cros01}), integrating over $\Omega_\pm$ and
$\phi_{\tau^-}$ angles, the spin-independent cross-section is
\begin{eqnarray}
\frac{d\: \sigma}{d\left(\cos\theta_{\tau^-}\right)}&&\left( e^+e^-
\stackrel{\tau^+\tau^-}{\rightarrow} h^+\bar{\nu}_\tau\,
h^-\nu_\tau\right)
=\frac{\pi\; \alpha^2}{2s}\beta\:  \left[(2-\beta^2\sin^2\theta_{\tau^-})\;
|F_1(s)|^2\right.\nonumber\\
&&\left.+4\, \, \re{F_2(s)}\right] Br(\tau^- \rightarrow
h^-\nu_\tau)\;Br(\tau^+
\rightarrow
h^+\bar{\nu}_\tau).
\label{eq:cros00}
\end{eqnarray}

By integrating over $\theta_{\tau^-}$, Eq.(\ref{eq:cros00}) will provide, to 
leading order in $\alpha$ (that means taking $F_1=1$, and $F_2=0$), the 
normalization for all the asymmetries considered in this paper.

This equation shows that, {\it a priori}, the real part of $F_2$ could be
measured from the cross section. However, this determination would be 
directly affected by the precision in the reconstruction of the Tau 
direction, so that
problems with dilution, kinematic reconstruction and/or efficiency over
$\cos\theta_{\tau^-}$, make difficult the extraction of $F_2$ from the
cross-section.

The
imaginary
part of $F_2$ is a $T$-odd, $C$- and $P$-even quantity; therefore, a suitable
observable to look for its determination will be the normal (to the scattering
plane) polarization of the outgoing $\tau$, as discussed after 
Eq.(\ref{cross3}). The polarization analyzer has been studied in detail in
Ref.~\cite{nos08} and the results are shown in Table 1.

In absence of electron
polarization, only spin-spin correlations of the Tau decay products allow
the determination of the real part of the $F_2$ magnetic moment form factor.
This will be the object of the following sections.

\section{Tau spin-spin correlations}\label{section:corr}

In this section we study several spin correlation observables proportional to
the magnetic form factor $F_2$ that could be measured at B-Factories. We will sum the events for all angles as possible in the angular distribution in such a way as to maintain the information on the magnetic form factor. In subsection 4.1
we first consider the case for direct Tau pair production, in subsection 4.2 we study the resonant Tau pair production observables and in subsection 4.3 we compute the contributions from $Z-\gamma$ interference.

\subsection{Direct production}

From Eq.(\ref{csection}), it is straightforward to see 
that the real part of the $F_2$ form factor can only be measured by
appropriate linear combinations of the different
spin correlation terms shown there, i.e, Transverse-Transverse ($TT\equiv xx$), Normal-Normal
($NN\equiv yy$), Longitudinal-Longitudinal ($LL\equiv zz$) and 
Longitudinal-Transverse ($LT\equiv zx$) to the scattering plane.

\subsubsection{With Tau-direction symmetric integration}

Integrating out $\cos\theta_{\tau^-}$ in Eq.(\ref{csection}) one gets that the
information carried by the LT correlation is erased, so that only $TT$, $NN$ and
$ZZ$ correlations survive

\bea
d^4\sigma^{SS}&=&\frac{2\pi\alpha^2\beta}{3\, s}\; \left[(s_+^x
s_-^x)\; \mathcal{XX} + (s_+^y s_-^y)\; \mathcal{YY}+ (s_+^z s_-^z)\;
\mathcal{ZZ}\right]\nonumber\\
&&\qquad\qquad\qquad\qquad \times\; \frac{d\Omega_{h^+}}{4\pi}\;
\frac{d\Omega_{h^-}}{4\pi}\; Br_+\;
Br_-
\label{eq:corr01}
\eea
being $Br_\pm$ the decay branching ratios of Taus to charged hadrons
plus neutrinos, $Br_-=Br(\tau^-\rightarrow h^-\nu_\tau)$,
$Br_+=Br(\tau^+\rightarrow  h^+\bar{\nu}_\tau)$, and
\bea
\mathcal{XX} &=&(2-\beta^2)\; |F_1|^2+4\re{F_2}\\
\mathcal{YY} &=& -|F_1|^2\, \beta^2\\
\mathcal{ZZ} &=&(1+\beta^2)\; |F_1|^2+2\re{F_2}
\eea

For disentangling the relevant $F_2$ form factor from the spin-spin correlations 
one may now integrate
appropriately the angular variables ($\Omega_\pm$) of the final hadrons. For 
each correlation, we show the way how this integration has to be done:
the $TT$ and $NN$ terms lead to azimuthal asymmetries of the decay products, whereas 
the $LL$ term leads to a polar asymmetry.

\begin{itemize}
\item For the $TT$ correlation, we integrate in Eq.(\ref{eq:corr01}) the polar
$\theta_\pm^*$ angles to get
\beq
\spa d^2\sigma_{TT}=\frac{\pi\alpha^2\beta}{96\, s}\;
\left[-(\alpha_-\alpha_+)\right]\; \left(\mathcal{XX}\right)\; 
\left(\cos\phi_-\; \cos\phi_+\right)\; d\phi_+\; d\phi_-\; Br_+\; Br_-
\eeq
and then perform an asymmetric integration over the azimuthal angles 
$\phi_\pm$. In this way we define the $TT$-asymmetry:
\bea
A_{TT}&\equiv&\frac{-(\alpha_-\alpha_+)}{\sigma}\;
\left(\int_{-\pi/2}^{\pi/2}d\phi_--\int_{\pi/2}^ {3\pi/2} d\phi_-
\right)\left(\int_{-\pi/2}^{\pi/2}d\phi_+-\int_{\pi/2}^{3\pi/2}d\phi_+
\right)\; d^2\sigma_{TT}\nonumber\\
&=&-\frac{\pi\alpha^2\beta}{6\, s}\;\frac{\alpha_-\alpha_+}{\sigma}
\left[(2-\beta^2)\; |F_1|^2+4\re{F_2}\right]\; Br_+\; Br_-\label{eq:TT}
\eea
with $\sigma$ (total cross section) given by the integration of
Eq.(\ref{eq:cros00}).
\item For the $NN$ term, we follow a similar procedure, integrating out the 
$\theta_\pm^*$ polar angles to get
\bea
d^2\sigma_{NN}&=&\frac{\pi\alpha^2\beta}{96\, s}\;
\left[-(\alpha_-\alpha_+)\right]\left(\mathcal{YY}\right)\;
\left(\sin\phi_-\; \sin\phi_+\right)\; d\phi_+\; d\phi_-\nonumber\\
&&\qquad\qquad\qquad\qquad\times\; Br_+\; Br_-
\eea
and then we integrate asymmetrically the azimuthal $\phi_\pm$ angles, in order to define to obtain the $NN$-asymmetry:
\bea
A_{NN}&\equiv&-\frac{\alpha_-\alpha_+}{\sigma}\;
\left(\int_{0}^{\pi}d\phi_--\int_{\pi}^ {2\pi} d\phi_-
\right)\left(\int_{0}^{\pi}d\phi_+-\int_{\pi}^{2\pi}d\phi_+
\right)\; d^2\sigma_{NN}\nonumber\\
&=&\frac{\pi\alpha^2\beta}{6\, s}\;\frac{(\alpha_-\alpha_+)}{\sigma}
\; \beta^2\; |F_1|^2\; Br_+\; Br_-\label{eq:NN}
\eea
\item Finally, for the $LL$ correlation we must first integrate the azimuthal
$\phi_\pm$ angles to get
\bea
\spa d^2\sigma_{LL}&=&\frac{\pi\alpha^2\beta}{6\, s}\;
\left[-(\alpha_-\alpha_+)\right]\; \left(\mathcal{ZZ}\right)\;
\left(\cos\theta_-^*\; \cos\theta_+^*\right)\; d(\cos\theta_+^*)\;
d(\cos\theta_-^*)\nonumber\\
 &&\qquad\qquad\qquad\qquad\qquad\times\; Br_+\; Br_-
\eea
and then integrate the polar angles $\theta_\pm^*$ asymmetrically, to finally define the $NN$-asymmetry as:
\bea
\spa A_{LL}&\equiv&-\frac{\alpha_-\alpha_+}{\sigma}\nonumber\\
&&\times\left(\int_{-1}^{0}d(\cos\theta_-^*)-\int_{0}^ {1} d(\cos\theta_-^*)
\right)\left(\int_{-1}^{0}d(\cos\theta_+^*)-\int_{0}^{1}d(\cos\theta_+^*)
\right)\; d^2\sigma_{LL}\nonumber\\
&=&-\frac{\pi\alpha^2\beta}{6\, s}\;\frac{(\alpha_-\alpha_+)}{\sigma}
\; \left[(1+\beta^2)\; |F_1|^2+2\re{F_2}\right]\; Br_+\; Br_-\label{eq:LL}
\eea
\end{itemize}

 \subsubsection{With Tau-direction asymmetric integration}
 
 In order to keep the information on $F_2$ from the $LT$ correlation 
as given by Eq.(\ref{eq:xz}) one must, contrary to the
previous cases, integrate first in an asymmetric form the $\theta_{\tau^-}$ 
angle of
the outgoing Tau
\bea
&&d^4\sigma^{SS}(FB)\equiv
\left(\int_{-1}^0d(\cos\theta_{\tau^-})-\int_{0}^1
d(\cos\theta_{\tau^-})\right)\frac{d\sigma^{SS}(\vec{n}^*_+,\vec{n}^*_-)}{d(\cos
\theta_{\tau^-})}\nonumber\\
&&\qquad\qquad\qquad\times \; \frac{d\Omega_{h^+}}{4\pi}\;
\frac{d\Omega_{h^-}}{4\pi}\; Br_+\; Br_-\nonumber\\
&&=\frac{2\pi\alpha^2\beta}{3\, s}\; 
\left[(n^*_+)^x (n^*_-)^z + (n^*_+)^z(n^*_-)^x\right]\; (\mathcal{ZX})\;
\frac{d\Omega_{h^+}}{4\pi}\;
\frac{d\Omega_{h^-}}{4\pi}\; Br_+\; Br_- 
\eea
with
\beq
\mathcal{ZX}=\frac{1}{\gamma}|F_1|^2+\gamma\; (2-\beta^2)\; \re{F_2}
\eeq
and 
\bea
&&(n^*_+)^x (n^*_-)^z +
(n^*_+)^z(n^*_-)^x=-\alpha_+\alpha_-\left(\sin\theta_-^*\cos\phi_-\cos\theta_+^*
\right.\nonumber\\
&&\qquad\qquad\left.
+(\phi_-\leftrightarrow\phi_+,\theta_-^*\leftrightarrow
\theta_+^*)\right)
\eea
Then, integrating out $\theta_-^*$ and $\phi_+$, and performing an asymmetric 
integration over
$\phi_-$, one can define a forward-backward $LT$-asymmetry as:
\bea
\spa d\sigma_{LT}&\equiv&-(\alpha_-\alpha_+)\nonumber\\
&\times&
\left(\int_{-\pi/2}^{\pi/2}d\phi_--\int_{\pi/2}^{3\pi/2} d\phi_-
\right)\; \left(\int_{0}^{2\pi}d\phi_+\right)\;\left(\int_{-1}^1 d(\cos\theta_-^*)\right)\; d^4\sigma^{SS}(FB)\nonumber\\
&=&-\frac{\pi\alpha^2\beta}{3\, s}\; (\alpha_-\alpha_+)\; \left(\mathcal{ZX}\right) \cos\theta^*_+\; d(\cos\theta_+^*)\; Br_+\; Br_-
\eea
Integrating now asymmetrically the $\theta_+^*$ angle, one gets the $LT$-Asymmetry
\bea
A_{LT}&\equiv&\left(\int_{-1}^0 d(\cos\theta_+^*)-\int_0^1 d(\cos\theta_+^*)\right)\;d\sigma_{LT}\nonumber\\
&=&\frac{\pi\alpha^2\beta}{6\, s}\;\frac{(\alpha_-\alpha_+)}{\sigma}
\; \left[\frac{1}{\gamma}|F_1|^2+\gamma\;(2-\beta^2)\;\re{F_2}\right]\; Br_+\; Br_-\label{eq:LT}
\eea

A similar procedure can be done by interchanging the angular variables of $h^+$ 
by those of $h^-$ in each of the previous steps. The result defines the 
$TL$-Asymmetry, which is numerically equal to the $LT$-Asymmetry
\beq
A_{TL}\equiv A_{LT}(+\leftrightarrow-)=A_{LT}
\eeq

\subsection{On top of the $\Upsilon$ resonances.}

As explained in the Introduction, our aim is to measure the observables on 
the top of the $\Upsilon$ peak where the $\tau$ pair-production is  mediated 
by the resonance. This has the advantage that resonant diagrams dominate the 
process so no contribution from box diagrams has to be considered. For Super 
B/Flavour factories, $\Upsilon(1S)$, $\Upsilon(2S)$ and $\Upsilon(3S)$ can be 
included since their decay rates into $\tau$ pairs have been measured and are sizeable.

The asymmetries obtained before are not modified when running on the top of a 
resonance. The differential cross section is only modified by a global 
$|H(M_\Upsilon^2)|^ 2$ factor, where the resonant amplitude is given by
\beq 
H(M_\Upsilon^2) =\frac{4\pi\alpha
Q^2_b}{M_\Upsilon^2}\frac{\left|F_\Upsilon\left(M_\Upsilon^2\right)\right|^2}
{i\Gamma_\Upsilon M_\Upsilon}=-i \,
\frac{3}{\alpha}
Br\left(\Upsilon \rightarrow e^+e^-\right)\,.
\label{factor}
\eeq
where $e-\tau$ universality has been assumed.

\subsection{ Contributions to the observables coming from the $Z-\gamma$ interference.}
 
On top of the $\Upsilon$ resonance, these contributions must, in principle, also 
be considered. This $Z$-mediated amplitude modifies the defined asymmetries as given by:
\bea
\spa A_{TT}^Z=-\frac{\pi\alpha^2\beta}{6\; M_\Upsilon^2}\;\frac{(\alpha_+\alpha_-)}{\sigma}\;(2-\beta^2)\; N_Z,\quad &&
A_{NN}^Z=\frac{\pi\alpha^2\beta}{6\; M_\Upsilon^2}\;\frac{(\alpha_+\alpha_-)}{\sigma}\; \beta^2\; N_Z,\label{eq:asimz1}\\
\spa A_{LL}^Z=-\frac{\pi\alpha^2\beta}{6\; M_\Upsilon^2}\;\frac{(\alpha_+\alpha_-)}{\sigma}\; (1+\beta^2)\; N_Z,\quad
&&
A_{LT}^Z=\frac{\pi\alpha^2\beta}{6\; M_\Upsilon^2}\; \frac{(\alpha_+\alpha_-)}{\sigma}\; \frac{1}{\gamma}\; N_Z\label{eq:asimz2}
\eea
with
\beq
N_Z=\frac{v\:v_b}{sw^2\: cw^2}\; \frac{Q_e}{Q_b}\; \frac{M_\Upsilon^2(M_\Upsilon^2-M_z^2)}{(M_\Upsilon^2-M_z^2)^2+\Gamma_z^2\: M_z^2}\; |H(M_\Upsilon^2)|^2
\eeq 
and 
\beq
a=-\frac{1}{2},\quad
v=-\frac{1}{2}+2s_w^2\quad v_b=-\frac{1}{2}+\frac{2}{3}s_w^2,\quad
Q_b=\frac{-1}{3},\quad Q_e=-1,
\eeq

Considering that the common numerical factor $N_Z$ is $(-1.567)\times 10^{-3}$, these 
$Z$-contributions to the asymmetries are not negligible and, in principle,  
have to be taken into account when considering the correlation observables. 
Fortunately, as can be seen from Eqs.(\ref{eq:asimz1},\ref{eq:asimz2}), because 
the $\gamma-\Upsilon-Z$ interference proceeds   
through the vector neutral current coupling to leptons, the relevant 
amplitudes contributing to the asymmetries have equal structure than
the ones obtained for the charge form factor $|F_1|^2$. As we will see in 
the following, the same combination of asymmetries which cancels the $F_1$ 
contribution will also cancel the contribution of the Z interference, so 
that it is possible to separate out $\re{F_2}$ from other contributions 
without any ambiguities.

\section{Precision on the measurement of $\re{F_2}$}

The  asymmetries $TT$, $NN$, $LL$ and $LT$ (Eqs.(\ref{eq:TT}), (\ref{eq:NN}), (\ref{eq:LL}) and 
(\ref{eq:LT}), respectively)  can be combined appropriately in order to eliminate the 
$|F_1|^2$ dependence. Then,
 the real part of the magnetic moment form 
factor can be obtained from the following three independent combinations 
of the four asymmetries:
\bea
\left(\begin{array}{l}
\null\\
\re{F_2}\\
\null
\end{array}\right)&=&\dps\frac{(3-\beta^2)}{(\alpha_+\alpha_-)\beta^2}\times \nonumber\\
&&\null\nonumber\\
&&\hspace*{-1cm}\left(\begin{array}{cccc}
\dps \frac{4}{\gamma^2\beta^2}&0&0&\dps\frac{4(2-\beta^2)}{\gamma\beta^2}\\[.2cm]
0&\dps \frac{4}{(3-\beta^2)\gamma^2}&0&\dps \frac{4(1+\beta^2)}{(3-\beta^2)\gamma}\\[.2cm]
0&0&\dps -\frac{4}{(2-\beta^2)\gamma^2}&\dps \frac{4\beta^2}{(2-\beta^2)\gamma}
\end{array}\right)\times
\left(\begin{array}{c}
A_{TT}\\
A_{LL}\\
A_{NN}\\
A_{TL}
\end{array}\right)
\label{eq:all}
\eea

Using these three independent combinations we can now estimate the precision 
that can be achieved on the determination of 
$\re{F_2}$. 

For our numerical analysis we assume the following  set of 
integrated luminosities: Babar +
Belle at $2  ab^{-1}$, and  a high statistics $B$/Flavour factory at 
$15 ab^{-1}$ (1 year running) 
and at $75 ab^{-1}$ (5 years running). The results are given in Table 1, where 
only the results for the $\pi^\pm$ ({\it i.e.} $h^\pm=\pi^\pm$ ) decay 
channel, for the traced $\tau^\pm$, are quoted. Results for the 
$\rho^\pm$ ($h^\pm=\rho^\pm$) and 
$\rho^\pm-\pi^\mp$ ($h^\pm=\rho^\pm, h^\mp=\pi^\mp$) channels are 4 and 2 
times looser, respectively,  
than those for the $\pi^\pm$ ones. They poorly contribute  to increase the 
precision measurement of $\re{F_2}$ but 
we have taken them into account, for completeness, in the global result given 
in Table 1. This global result 
has been obtained by considering all correlation channels of Eq.(\ref{eq:all}), 
for all possible combination of the  Tau decays to $\pi$'s and $\rho$'s. For all 
the results presented here only statistical errors are considered. The 
comparison with the results given in Ref.~\cite{nos08} shows that observables 
obtained with polarized electrons are better by a factor 3 than those given 
for unpolarized electron-beams.   
In addition, one must be aware that experimental uncertainties may be bigger 
for the detection in coincidence of two Taus, in order to get correlations, than 
for single Tau polarization, so that bounds from correlations may be looser 
that those for single Tau observables.

\def\up{-7pt}
\def\tsize{\normalsize}
\def\sm{\small}
\def\ft{\footnotesize}
\begin{center}
\begin{table}[hbt]{\centering
\caption{Sensitivity of the $F_2(q^2)$ form factor measurement at the $\Upsilon$ energy}
\begin{tabular}{|c|c||c|c|c|}
\hline
\multicolumn{2}{|c||}{\null}&\multicolumn{3}{|c|}{ E X P E R I M E N T ($ab\,
=\, {\rm atto barn}\,=\,10^{-18}b$)}\\
\cline{3-5}
\multicolumn{2}{|c||}{\null}& &\multicolumn{2}{|c|}{Super B/Flavour Factory}\\ \cline{4-5}
\multicolumn{2}{|c||}{\null}& Babar+Belle &(1 yr. running)&(5 yrs. running)\\ \cline{1-2}
&Correlations&$2ab^{-1}$ &$15ab^{-1}$ &$75ab^{-1}$ \\ \cline{2-5}
&$TT--LT$ & $7.6\times 10^{-5}$ & $2.8\times 10^{-5}$ & $1.2\times 10^{-5}$\\ \cline{2-5}
$\re{F_2}$&$LL--LT$ & $5.2\times 10^{-5}$ & $1.9\times 10^{-5}$ & $8.5\times 10^{-6}$\\ \cline{2-5}
&$NN--LT$ & $5.1\times 10^{-5}$ & $1.8\times 10^{-5}$ & $8.3\times 10^{-6}$\\ \cline{2-5}
&Global & $2.9\times 10^{-5}$ & $1.1\times 10^{-5}$ & $4.7\times 10^{-6}$\\ \hline\hline
$\begin{array}{c} \im{F_2}\\[-.2cm]
\mbox{\scriptsize (from}\\[-.2cm]
\mbox{\scriptsize Ref.\cite{nos08})}\end{array}$ 
&$\begin{array}{c}\mbox{Normal}\\[-.2cm]
\mbox{single-}\tau\\[-.2cm]
\mbox{Asymm.}\end{array}$& $2.1\times 10^{-5}$ & $7.8\times 10^{-6}$ & $3.5\times 10^{-6}$\\ \hline
\end{tabular}}
\label{table}
\end{table}
\end{center}

\section{Conclusions}

We have estimated the precision that can be achieved in 
the determination of the $QED$ scale and flavour effects in the 
measurement of the $F_2$ magnetic moment form factor, at 
$\Upsilon$ energies, for unpolarized $e^+\, e^-$ collisions. 
In Ref.~\cite{nos08} we already showed that the imaginary part of $F_2$ can be determined from the Normal Asymmetry of the 
decay products for a normal polarization of a single Tau. In this paper, we have shown that, for
 unpolarized electron beams, the real part of the form factor needs the measurement 
 of correlations on the Tau decay products of both polarized Taus. We have defined 
 correlations sensitive to the $F_2$ form factor and found three independent 
 combinations of them to get its value without any ambiguities, eliminating 
 the contribution of the charge form factor $F_1$ and the $Z-\gamma$ interference. 
 Combining all channels, the sensitivity that can be achieved is of the order 
 of $10^{-5}-10^{-6}$, which allows the measurement of the magnetic moment 
 form factor $F_2(M_\Upsilon^2)$, given in Eq.(\ref{f2}), up to a precision of a 
 few per cent. This result shows that Super B/Flavour factories can bring, for the first time, 
 important information on the rather poorly known magnetic properties of the Tau.

\begin{ack}
This work has been supported by the Ministerio de Educaci\'on y Ciencia (MEC) and FEDER, under the grants FPA2005-00711 and FPA2005-01678, and by Pedeciba and DINACYT-\-PDT-\-54/94-Uruguay.
\end{ack}

\end{document}